\pdfoutput=1 
\documentclass{aipproc}
\usepackage{setspace}
\usepackage[all]{xy}
\usepackage{graphicx, color}
\usepackage{float, subfigure, caption}
\usepackage{amsmath, amsthm, amssymb, amscd, mathrsfs}
\usepackage{psfrag}

\newcommand\selectedlayoutstyle {8x11single}
\layoutstyle\selectedlayoutstyle

\SetInternalRegister\hbadness{8000} 

\begin{document}

\newcommand{\alice}{\textsc{ALICE}}
\newcommand{\colstar}{\textsc{STAR}}
\newcommand{\herwig}{\textsc{HERWIG}}
\newcommand{\pythia}{\textsc{PYTHIA6}}
\newcommand{\qpythia}{\textsc{QPYTHIA}}
\newcommand{\fastjet}{\textsc{\-FASTJET}}
\newcommand{\embedding}{\emph{embedding}}
\newcommand{\versus}{\emph{versus}}
\newcommand{\IPE}{\emph{Independent Particle Emission}}

\newcommand{\dpT}{$\delta p_{T}$}
\newcommand{\Ajato}{$A_\text{jato}$}

\newcommand{\note}[1]{\textbf{(\textcolor{red}{#1})}}
\newcommand{\topic}[1]{\textbf{$\bullet$ #1}}

\title
      [Inclusive Distribution of Fully Reconstructed Jets in Heavy Ion Collisions at RHIC: Status Report]
      {Inclusive Distribution of Fully Reconstructed Jets in Heavy Ion Collisions at RHIC: Status Report}


\author{G.~O.~V.~de~Barros\thanks{gbarros@dfn.if.usp.br} ~for the STAR Collaboration}
{
  address={Instituto de Física, Universidade de São Paulo, Brazil}
}

\begin{abstract}
We present an update to the analysis of fully reconstructed jets in
heavy ion collisions by the STAR  Collaboration at RHIC. We
analyse the response of the anti-$k_{T}$ algorithm in
the presence of background, and present a new observable for the measurement
of inclusive jet production that is expected to be more robust against
background model assumptions than previous jet analyses at RHIC and LHC.
\end{abstract}

\date{\today}

\maketitle

\section{Introduction}

The interaction of a hard-scattered parton with a colored medium, known as jet
quenching, is a key probe of the Quark-Gluon Plasma (QGP) generated in
high energy heavy ion collisions. In a QCD picture, jet quenching is
dominated by gluon bremsstrahlung, leading to softening of the
typical jet fragmentation pattern relative to jets in vacuum. The
imposition of kinematic cuts on the jet constituents, including
``jet'' measurements via single leading hadrons, results in an
apparent medium-induced energy loss of the jet, though full
reconstruction of the jet should recover its full energy. Jets of finite 
acceptance (finite R) may nevertheless not recover the full jet energy due to
large angle radiation; indeed, that is expected to be the clearest way to 
measure the phenomenon.

The most direct measurement of jet production and apparent energy loss due to
quenching is the inclusive jet production. Such jet measurements are
challenging in the complex environment of heavy ion collisions, however, due to
the very large flux of soft hadrons from other, incoherent interactions in the
collision. The STAR Collaboration at RHIC has previously presented the analysis
of inclusive jet and hadron-jet coincidence rates in central $Au+Au$ collisions
\cite{bib:SevilJets,bib:mploskon_qm09}, in which minimal kinematic cuts on the
jet constituents are imposed, to achieve minimal bias of the resulting 
distributions. However, these analyses employed model assumptions for the 
influence of background on the jet measurements, and imposed kinematic cuts on
the measured (i.e. jet+background) distributions themselves. Both effects need
careful assessment to establish their systematic uncertainties.

In these proceedings we present a new, data-driven approach, in which
the full spectrum of reconstructed jets is analysed in order to
measure the fluctuations of the underlying event. We then introduce a
new, differential, observable sensitive to the rate of inclusive jet
production in nuclear collisions, in which the relative background
contribution is substantially reduced relative to previous
measurements, and indicate methods to assess the remaining background
quantitatively.

\section{Data sample and Jet Reconstruction}
This study uses $\sim 1M$ central ($0-10\%$) $Au+Au$ events at
 $\sqrt{s_{NN}} = 200\text{ GeV}$ collected by STAR during RHIC run 7. Jet 
reconstruction utilizes charged particles measured in the Time Projection 
Chamber (TPC) and neutral energy measured in the Barrel Electromagnetic 
Calorimeter (BEMC) \cite{bib:STARSpinJets}. All tracks and towers with
$p_{T} > 0.2\text{ GeV/c}$ are used as input for the jet reconstruction. Jets
are reconstructed by $k_{T}$ \cite{bib:kt} and anti-$k_{T}$\cite{bib:anti-kt}
algorithms with resolution parameter $R=0.4$ using an energy recombination 
scheme, as implemented in \fastjet\, \cite{bib:fastjet}. The anti-$k_{T}$
algorithm is used to reconstruct signal jets and the $k_{T}$ algorithm to
evaluate the median background energy density. A jet is accepted if its
centroid lies within $|\eta| < 0.6$.

Because of the complex interplay between the hard jet component (high $Q^{2}$)
and fluctuations in measured jet energy due to the underlying event, in
this analysis we define a ``jet'' operationally, simply as the direct output
of the reconstruction algorithm. We relegate the separation of hard
jet and background to a separate step. The ``Direct'' jet transverse
energy is defined as \cite{bib:bkg_subtraction}:

\begin{equation}
  p_{T}^\text{direct}= p_{T}^\text{reco} - \rho\cdot A_\text{reco jet},
  \label{eq:pTdirect}
\end{equation}

\noindent
where $p_{T}^\text{reco}$ is the reconstructed jet transverse energy (anti-$k_{T}$), $A_\text{reco jet}$ is
its area \cite{bib:jet_area}, and
$\rho = \text{median}\left\{p_{T}^\text{reco,i}/A_\text{reco jet,i}\right\}$
is the event-wise median energy density evaluated with $k_{T}$ algorithm
\cite{bib:bkg_subtraction}. Therefore, $p_{T}^\text{direct}$ is corrected on an
event-wise basis for the overall level of background, but not for its 
region-to-region fluctuations.

In this study we consider only jets with $A_\text{reco jet}>0.4sr$, which is
efficient for hard jets in large background but biases against jets
comprising soft background only \cite{bib:pmj_hp10}.

\section{Analysis and Results}

Jet quenching corresponds to the medium-induced modification of jet
fragmentation, and robust reconstruction of quenched jets should have
minimal sensitivity to details of the fragmentation. In order to
quantify the sensitivity of anti-$k_{T}$ reconstruction to
fragmentation in the presence of large
background, we embed simulated jets with known structure into real
STAR central $Au+Au$ events, apply jet reconstruction to the hybrid events, identify
the reconstructed jet containing the embedded objection, and calculate \cite{bib:pmj_hp10}:

\begin{equation}
  \delta p_{T} = p_{T}^\text{reco} - \rho\cdot A_\text{reco jet} - p_{T}^\text{embed},
  \label{eq:dpT}
\end{equation}
\noindent
where $p_{T}^{embed}$ is the embedded transverse energy. 

\begin{figure}[h]
  \centering
  \includegraphics[width=.52\textwidth]{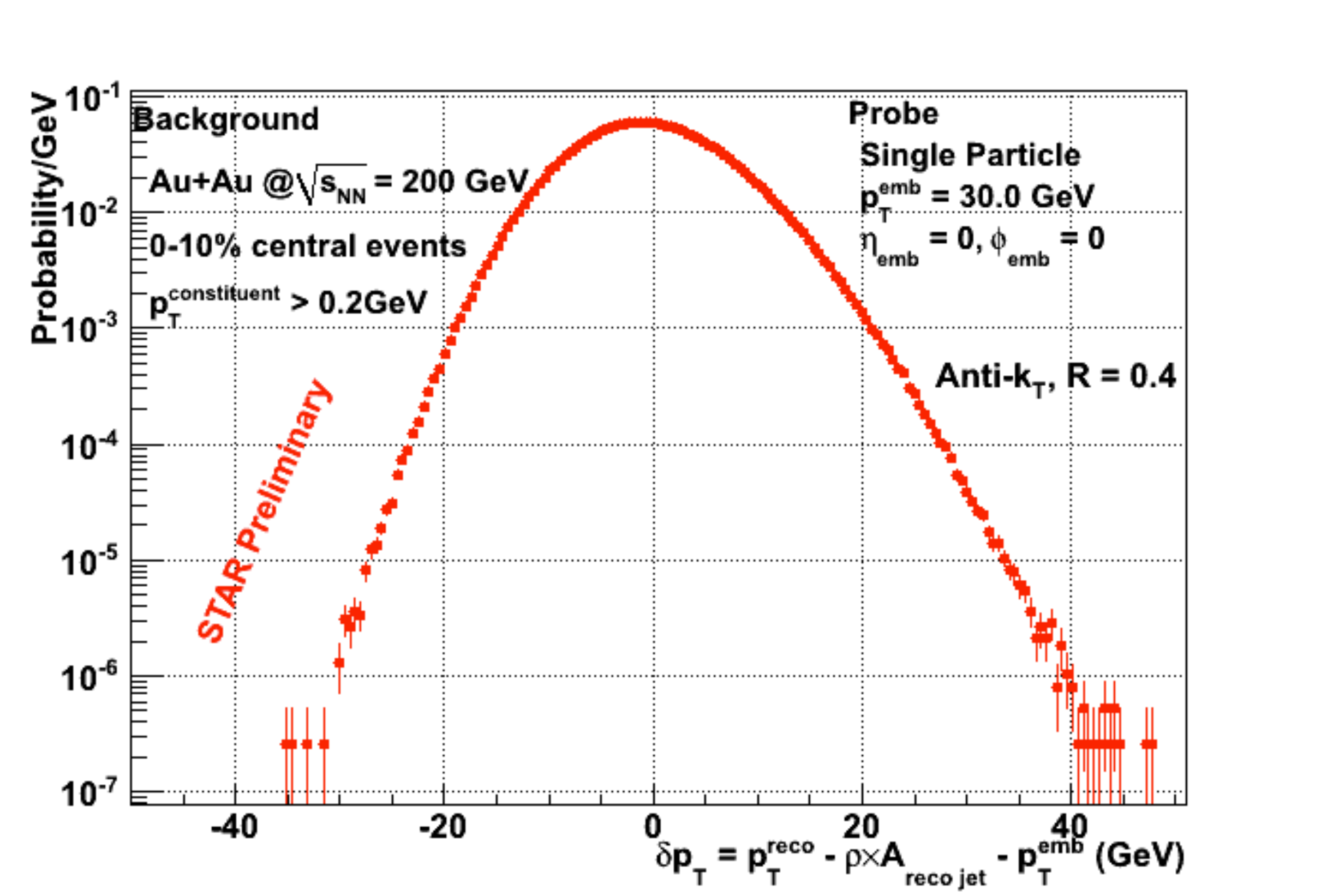}
  \includegraphics[width=.52\textwidth]{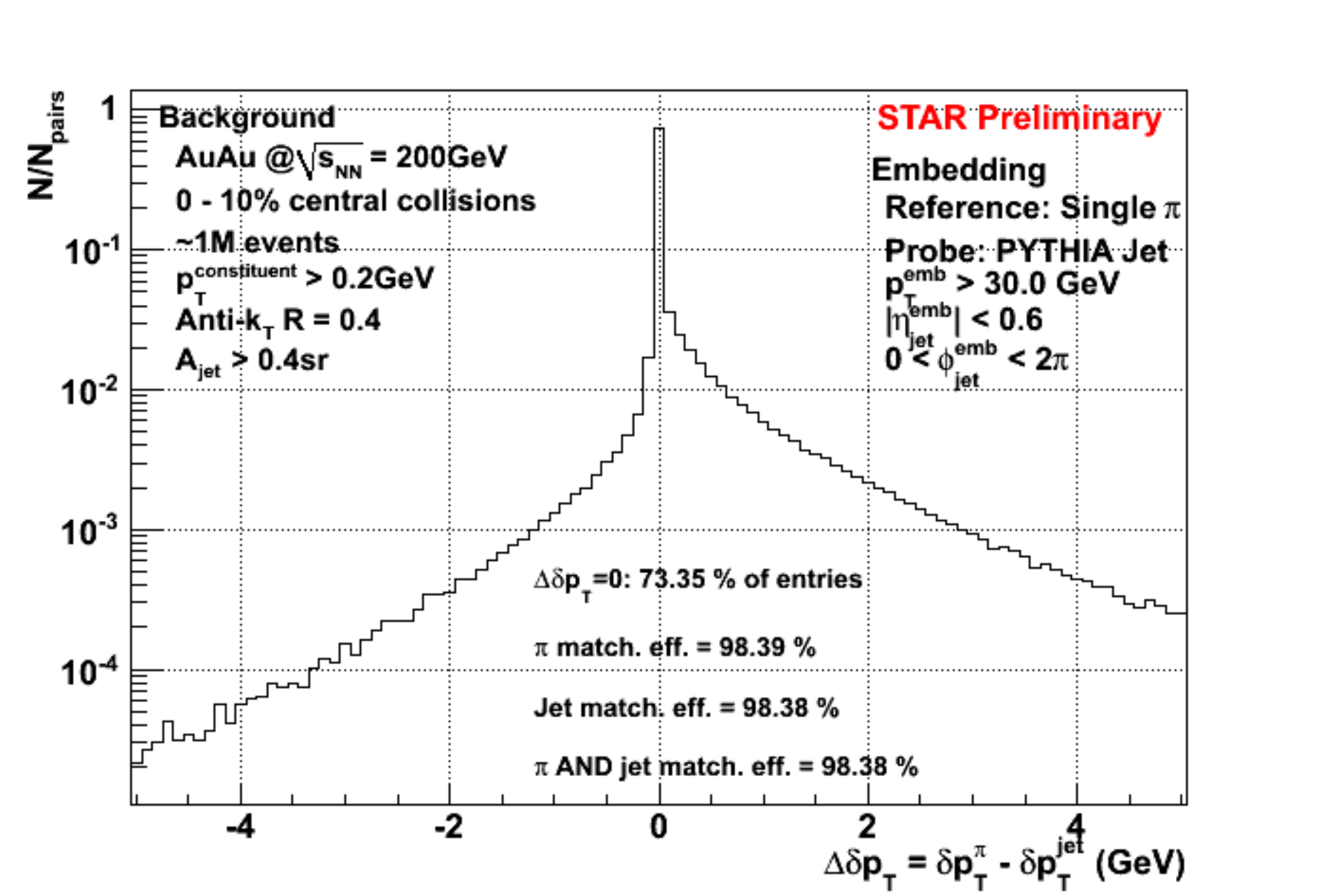}
  \caption{Left: response of jet reconstruction for a single pion
    embedded with $p_{T} = 30\text{ GeV}$. Right: event-wise comparison
    of response (eq. \eqref{eq:dpT}) to a PYTHIA embedded jet $(p_{T} >
    30\text{ GeV})$ and a single pion with same kinematics.}
  \label{fig:fragmentation_pattern}
\end{figure}

We have explored a variety of jet fragmentation patterns and jet energies:
single pions, and jets generated at $\sqrt{s} = 200\text{ GeV}$ via
PYTHIA (p+p) and QPYTHIA (``p+p quenched'') \cite{bib:QPYTHIA}.
Figure \ref{fig:fragmentation_pattern}, left panel, shows the response
$\delta p_{T}^{\pi}$ for single 30 GeV pions averaged over 8M
different STAR central events \cite{bib:pmj_hp10}. To look in more
detail, the specific sensitivity to fragmentation is isolated by
comparing $\delta p_{T}$ event-wise for two different fragmentation
patterns embedded into the same event.  Figure
\ref{fig:fragmentation_pattern}, right panel, shows the distribution
in event-wise difference
\begin{equation}
  \Delta\delta p_{T} = \delta p_{T}^{\pi} - \delta p_{T}^\text{jet},
  \label{eq:Ddpt}
\end{equation}
between $\delta p_{T}$ for a PYTHIA-generated jet
with $p_T>30$ GeV (``probe'') and that for a single pion with the
same $p_T,\eta,\phi$. The distribution is calculated using 1M
different PYTHIA jets/single pions in different STAR events. With high
probability ($\sim73\%$ in this case), the response of anti-$k_T$ is
within 200 MeV the same for two very different fragmentation
patterns. Similar results are obtained for much lower jet $p_T$, and
for QPYTHIA-generated quenched jets. This quantifies the {\it insensitivity} of
anti-$k_T$ to the fragmentation pattern in the presence of a heavy-ion 
background event, a crucial property for jet quenching studies. Anti-$k_T$
appears to respond geometrically \cite{bib:anti-kt}.

\begin{figure}[h]
  \centering
  \includegraphics[width=.52\textwidth]{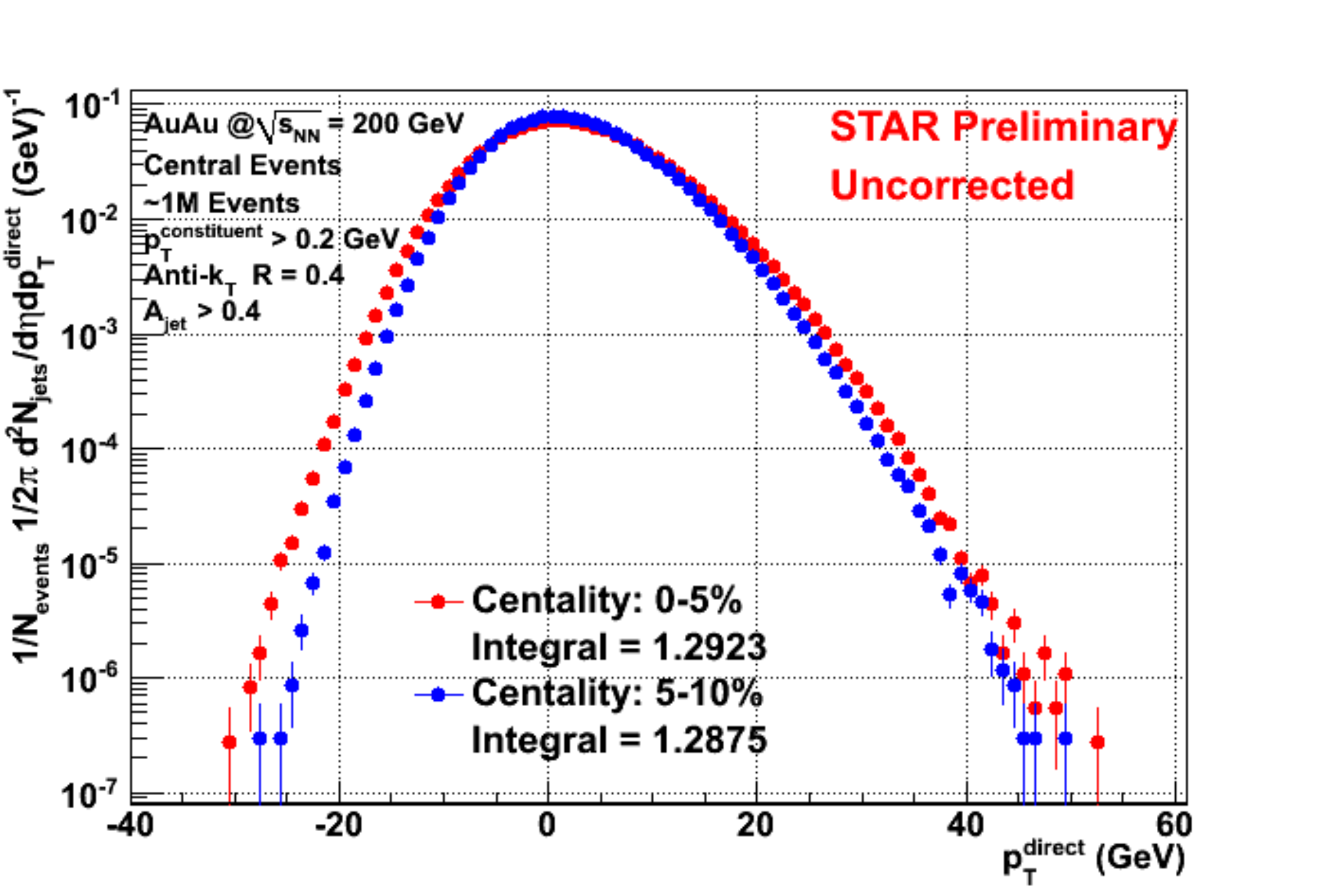}
  \includegraphics[width=.52\textwidth]{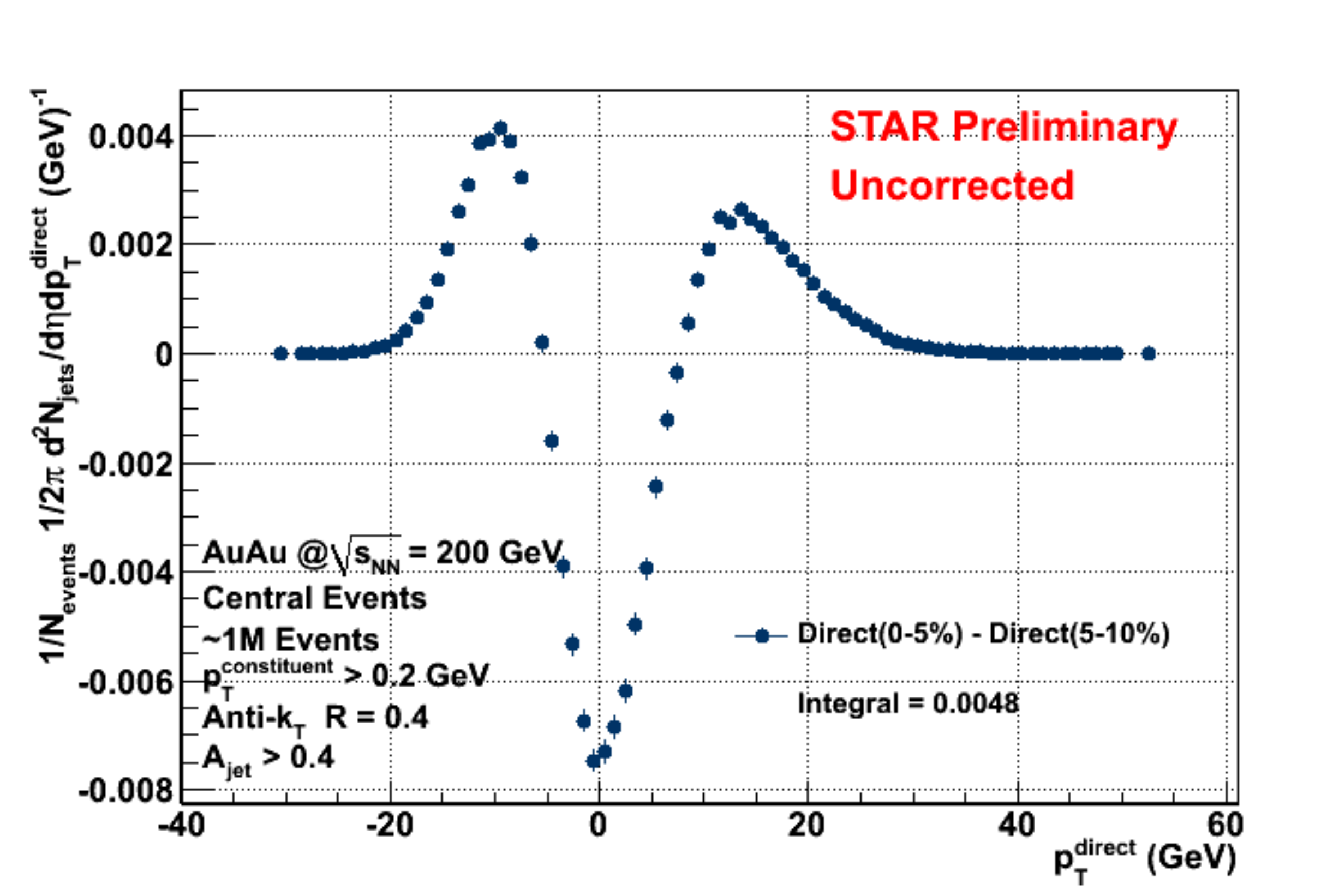}
  \caption{(Color online) Left: Direct jet spectra for two different
    event centralities: $0-5\%$ (red) and $5-10\%$ (blue). Right:
    difference between the two spectra.}
  \label{fig:jet_spectrum}
\end{figure}

Figure \ref{fig:jet_spectrum}, left panel, shows the Direct jet energy
distribution (eq. \eqref{eq:pTdirect}) for two different centrality classes: 
$0-5\%$ and $5-10\%$. Due to the operational definition in this
analysis of a jet simply as the output of the jet reconstruction corrected 
event-wise for median background level, a large fraction of the ``jets'' in the
distribution have apparently unphysical negative energy. This is
trivially due to the fact that $\rho$ is the median energy density. However, we
expect the negative energy ``jets'' to be dominated by combinatorial
jets arising from incoherent soft processes, and will use their distribution
to constrain our estimate of background fluctuations. In addition, the hard jet
component lies on top of the fluctuations and is broadly distributed in the 
figure; imposing any cut on $p_{T}^\text{direct}$ introduces an arbitrary
threshold whose effect is difficult to quantify.

The centrality classes in Figure \ref{fig:jet_spectrum}, left panel,
differ in average charged particle multiplicity by $\sim 12\%$. 
In contrast, we find that the reconstructed jet multiplicity in the acceptance 
for $A_\text{reco jet}>0.4sr$ differs between the two classes by less than
$\sim 0.3\%$. We have found that this approximate invariance of jet 
multiplicity persists over at least a factor two variation in charged
multiplicity. This again indicates that the response of anti-$k_{T}$ is
dominantly geometric: the number of jets in the acceptance is to a large extent
constant, while the Direct energy distribution varies as the relative hard and 
combinatorial components of the event change.

This observation suggests that the hard jet component might be best
isolated by observing the {\it evolution} of the Direct jet spectrum
with change in event structure, specifically with change in
centrality. Figure \ref{fig:jet_spectrum}, right panel, shows the {\it
  difference} between the two distributions in the left panel,
illustrating the redistribution of Direct jet energy with change in
centrality: for more central events there is an enhancement in both
background $(p_{T}^\text{direct} \lesssim -10\text{ GeV})$ and signal
$(p_{T}^\text{direct} \gtrsim 10\text{ GeV})$ regions. The dip about
zero is another manifestation of the observed conservation of total jet
multiplicity.

We now utilize the region $p_{T}^\text{direct} < 0$ to give a data-driven
estimate of the combinatorial (background) jet distribution, by
fitting a Gamma function in the negative region and extrapolating to the 
positive region. This functional form for the distribution of transverse energy
in a finite acceptance corresponds to the independent emission of a large
number of particles drawn from an exponential $p_T$ distribution
\cite{bib:tannenbaum}. A related function has been found
to describe the distribution of a closely related observable, event-wise mean
$p_T$ $\left(\langle p_{T}\rangle\right)$ at the SPS \cite{bib:SPSMeanpT} and
RHIC \cite{bib:RHICMeanpT, bib:RHICpTcorr}. The distribution has contributions
from multiple sources of fluctuations, both few-particle and global, such as
reaction plane orientation and residual dependence on event centrality. The 
Gamma-fn. parametrization includes all such effects with proper averaging.

\begin{figure}[h]
  \centering
  \includegraphics[width=.52\textwidth]{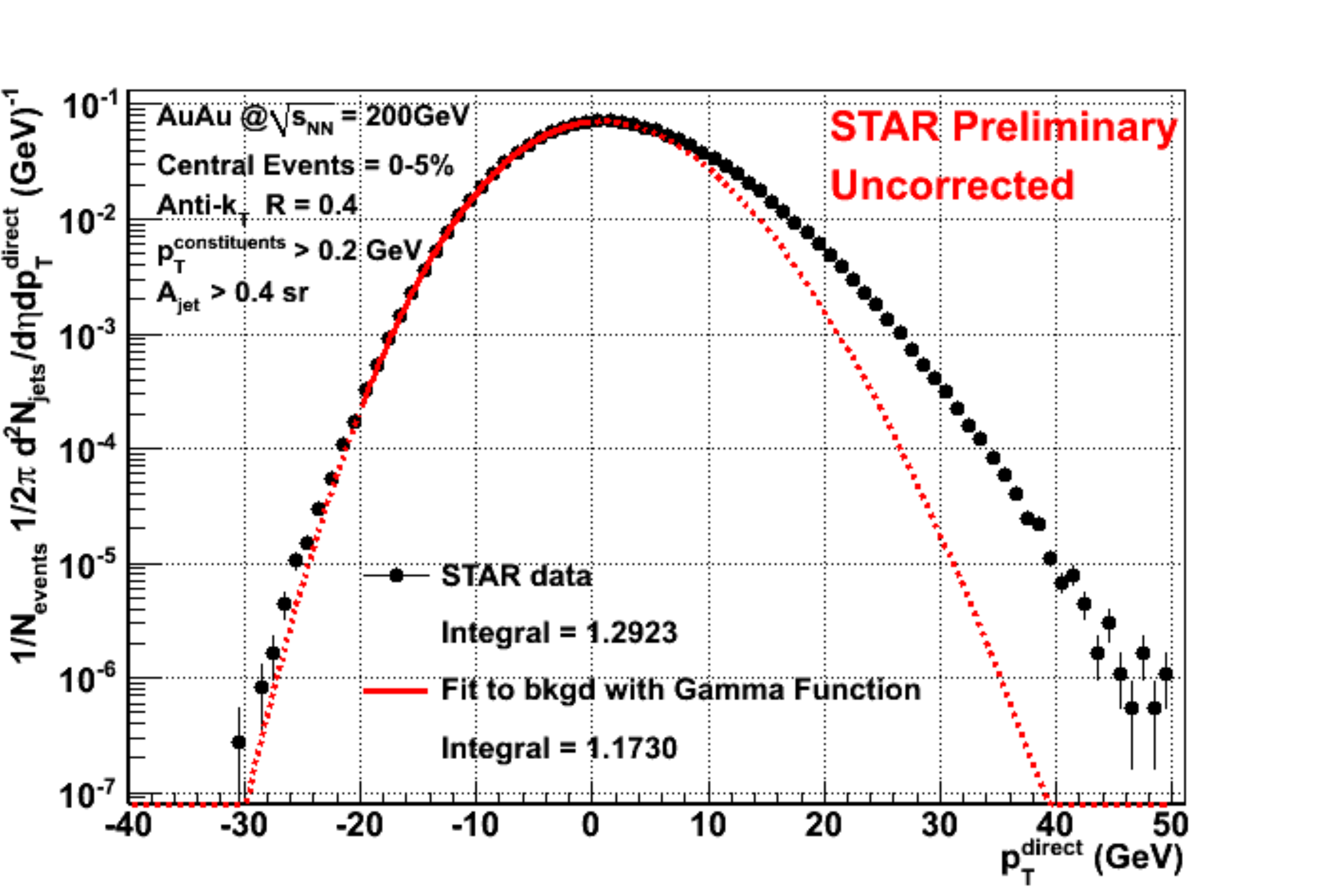}
  \includegraphics[width=.52\textwidth]{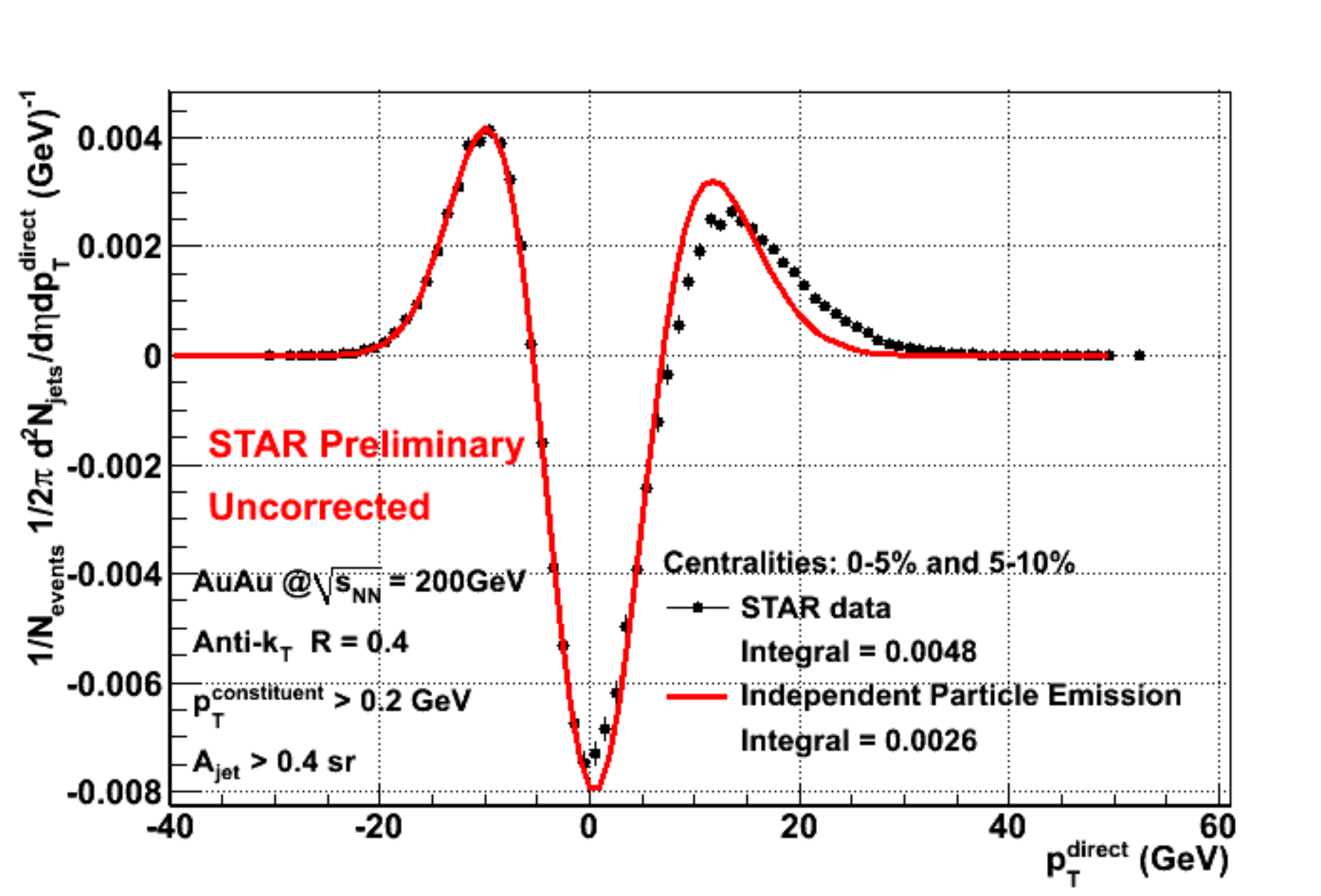}
  \caption{Direct jet spectrum for $0-5\%$, fitting and extrapolation (left
    panel) and difference spectrum and difference between fitted functions 
    (right panel).}
  \label{fig:gammafits}
\end{figure}

Figure \ref{fig:gammafits}, left panel, shows the Gamma-fn. fit to the
negative region, and extrapolation to the positive region, for the Direct jet
spectrum for $0-5\%$ central events. The fit describes the negative
region well over $\sim3-4$ orders of magnitude. The right panel
compares the difference between the fits for different centralities to
that for the data (same as Fig. \ref{fig:jet_spectrum}, right panel). As
expected, there is a good agreement between the data and the fit in
the negative region of the difference spectrum. The fit difference
overshoots the data difference at moderate $p_{T}^\text{direct} > 0$
while undershooting it for $p_{T}^\text{direct} \gtrsim
10\text{ GeV}$. This is expected, since the growth at large 
$p_{T}^\text{direct}$ for more central collisions is due in part to additional 
hard jets, which must displace a fraction of soft jets in order to conserve 
total jet number.

\section{Outlook}

The final analysis of the hard jet component is beyond the scope of these
proceedings. Further systematic investigation of background corrections, that 
are independent of the Gamma-function ansatz, are in progress. Alternative
functional forms for the background estimation are also under investigation.
However, we note that an important check on the hard component to be extracted 
from Fig. \ref{fig:gammafits}, right panel, comes from the well-established
binary-collision scaling of hard process rates in nuclear
collisions. Glauber modeling indicates that $<N_{bin}>\sim1000$ for
the 0-5\% centrality class, while $<N_{bin}>\sim800$ for 5-10\%. Thus,
difference spectrum in Fig. \ref{fig:jet_spectrum} must reflect the
jet production rate due to $\sim200$ additional binary
collisions. Implicit, of course, is the assumption that quenching
effects do not vary significantly between the centrality bins.

\bibliographystyle{aipproc}
\bibliography{bibliography}

\end{document}